\documentstyle[12pt,aasms4,graphics]{article}
\newcommand{\etal}{et~al.~}
\def\simlt{\mathrel{\hbox{\rlap{\hbox{\lower4pt\hbox{$\sim$}}}\hbox{$<$}}}}
\def\simgt{\mathrel{\hbox{\rlap{\hbox{\lower4pt\hbox{$\sim$}}}\hbox{$>$}}}}

\def\ale{\mathrel{\hbox{\rlap{\hbox{\lower4pt\hbox{$\sim$}}}\hbox{$<$}}}}
\def\age{\mathrel{\hbox{\rlap{\hbox{\lower4pt\hbox{$\sim$}}}\hbox{$>$}}}}
\def\farcs{\hbox{$.\!\!^{\prime\prime}$}}

\begin{document}
\oddsidemargin=0mm

\title{
On the Progenitors of Two Type II-P Supernovae in the Virgo Cluster
\footnote{
Based on observations made with the NASA/ESA {\it Hubble Space Telescope},
obtained at the Space Telescope Science Institute, which is operated
by the Association of Universities for Research in Astronomy, Inc.,
under NASA contract NAS 5-26555. These observations are associated with 
program GO-10877.}
\footnote{
Based on observations obtained with MegaPrime/MegaCam, a joint project of CFHT
and CEA/DAPNIA, at the Canada-France-Hawaii Telescope (CFHT) which is operated
by the National Research Council (NRC) of Canada, the Institute National des
Sciences de l'Univers of the Centre National de la Recherche Scientifique
of France, and the University of Hawaii.
}
}

\author{Weidong Li\altaffilmark{3}, Xiaofeng Wang\altaffilmark{3,4},
Schuyler D. Van Dyk\altaffilmark{5}, Jean-Charles Cuillandre\altaffilmark{6},
Ryan J. Foley\altaffilmark{3}, and Alexei V. Filippenko\altaffilmark{3} \\
Email: (wli, wangxf, rfoley, alex)@astro.berkeley.edu, vandyk@ipac.caltech.edu, jcc@cfht.hawaii.edu}

\altaffiltext{3}{Department of Astronomy, University of California,
Berkeley, CA 94720-3411.}

\altaffiltext{4}{Tsinghua Center for Astrophysics (THCA) and Physics
Department, Tsinghua University, Beijing, 100084, China.}

\altaffiltext{5}{Spitzer Science Center, California Institute of
Technology, Mailcode 220-6, Pasadena, CA 91125.}

\altaffiltext{6}{Canada-France-Hawaii Telescope Corporation, 65-1238
Mamalahoa Hwy, Kamuela, HI 96743.}

\slugcomment{Accepted by ApJ}

\begin{abstract}

Direct identification of the progenitors of supernovae (SNe) is rare
because of the required spatial resolution and depth of the archival data
prior to the SN explosions. Here we report on the identification of the
progenitors of two nearby SNe in the Virgo cluster: SN 2006my in NGC
4651 and SN 2006ov in M61. We obtained high-quality ground-based
images of SN 2006my with the Canada-France-Hawaii Telescope, and are
able to locate the site of the SN on pre-SN {\it Hubble Space
Telescope\/} ({\it HST}) Wide Field Planetary Camera 2 images
to a high precision (1$\sigma$ uncertainty of $\pm0\farcs$05).  We
pinpoint the site of SN 2006ov to within 0$\farcs$02 from {\it HST}
Advanced Camera for Surveys images of the SN.  We detected a red
supergiant progenitor for each SN within the error circles, with an
inferred zero-age main-sequence mass ($M_{\rm ZAMS}$) of $10^{+5}_{-3}$ $M_\odot$ and 
$15^{+5}_{-3}$
$M_\odot$ for the progenitors of SNe 2006my and 2006ov, respectively.
The mass estimates for the progenitors of both SNe confirm a suggested
trend that the most common Type II-plateau SNe originate from low-mass
supergiants with $M_{\rm ZAMS} \approx 8$--20~$M_\odot$.

\end{abstract}


\section{Introduction}

When a bright, nearby supernova (SN) is discovered, it may be 
possible to directly identify the progenitor of the SN in deep pre-SN
images. Analysis of the properties (e.g., mass, spectral type) of such
SN progenitors allows a direct comparison to theoretical expectations
based on the observed quantities of the SNe (e.g., light curves,
spectral evolution).

While observers have had some luck with ground-based observations,
such as the identifications of a $\sim20~M_\odot$ blue supergiant
(BSG) star for the peculiar, subluminous Type II SN 1987A in the Large
Magellanic Cloud (Gilmozzi \etal 1987; Sonneborn \etal 1987), a 
$\sim17~M_\odot$ red supergiant (RSG), possibly in a binary system, for the
Type IIb SN 1993J in M81 (Aldering \etal 1994; Van Dyk \etal 2002),
and a 13--20~$M_\odot$ yellow supergiant (YSG) for the Type II-P SN
2004et in NGC 6946 (Li \etal 2005a, 2005b), most of the recent
progress in this field has been based on the rich archival data of the
{\it Hubble Space Telescope} ({\it HST}) (Van Dyk \etal 2003a; Smartt
\etal 2004; Maund \& Smartt 2005; Maund \etal 2005a; Li et al. 2006;
Hendry \etal 2006; Gal-Yam \etal 2007).  The superior spatial
resolution of the {\it HST} images allows the SN progenitors to be
isolated from their environments, and the deep limiting magnitude
allows the SN progenitors to be detected in more distant (but still
relatively nearby) galaxies.

In this paper we report the identification of the progenitors on {\it
HST} images of two nearby SNe in the Virgo cluster.  K. Itagaki, a
veteran Japanese amateur SN searcher, discovered both of these SNe: SN
2006my in NGC 4651 on 2006 Nov. 8.82 (UT dates are used throughout this
paper) and SN 2006ov in M61 (NGC 4303) on 2006 Nov. 24.86 (Nakano \&
Itagaki 2006a, 2006b).  Unfortunately, the SN explosions occurred
while the galaxies were behind the Sun, and were only detected after
becoming visible again in the early morning, so the objects were
discovered rather late in their evolution.  Spectra obtained soon
after the time of discovery show that both SNe are of Type II: SN
2006my is similar to SN II-P 1999em (Hamuy \etal 2001; Leonard \etal
2002b; Elmhamdi \etal 2003) about 1--2 months after maximum 
brightness (Stanishev \&
Nielsen 2006), and SN 2006ov, perhaps reddened, is similar to SN II-P
2005cs (Pastorello \etal 2006; Tak\'ats \& Vink\'o 2006) roughly a
month after maximum (Blondin \etal 2006) (see \S 2 for a more detailed
discussion of the age and reddening of both SNe).

We took ground-based images of the SNe with the Canada-France-Hawaii
Telescope (CFHT) under good seeing conditions (Figures 1 and 2), 
and are able to
precisely (to within $\sim 0\farcs05$ and $0\farcs11$, for SNe 2006my
and 2006ov, respectively) locate the site of both SNe on
pre-SN archival {\it HST} images of the host galaxies. We further
obtained {\it HST}/ACS images of SN 2006ov itself and were able to
pinpoint the SN site to within 0$\farcs$02. We identify a 
7--15~$M_\odot$ red supergiant (RSG) and a 12--20 $M_\odot$ RSG in the error
circles of SNe 2006my and 2006ov, respectively, and propose these as
the progenitors of the SNe. We note that NGC 4651 also produced SN 1987K, 
while M61 produced SNe 1926A, 1961I, 1964F, and 1999gn. 

This paper is organized as follows. In \S 2 we discuss the properties
(type, age, and reddening) of both SNe, and demonstrate that they are of
the Type II variety discovered near the end of the plateau phase.  In
\S 3 we report our identification of the progenitors in the pre-SN
{\it HST}/WFPC2 images. We describe the nature of the progenitors
inferred from the {\it HST} photometry in \S 4. Further discussion is
in \S 5, and we summarize our conclusions in \S 6.

\section{Observations of SN 2006my and SN 2006ov}

There are several subclasses of SNe~II (see Filippenko 1997 for a
review of SN types). Among these, the Type II-plateau (SNe~II-P) are
the most common, with the defining feature being a prominent plateau
phase in their optical light curve. The Type II-linear (SNe~II-L) show
a linear decline in magnitude after their maximum brightness, and are
relatively rare.  The SN 1987A-like objects are subluminous and show a
very broad peak in the light curve. The Type II-narrow (SNe~IIn),
which have relatively narrow spectral features sometimes (but not always)
superimposed on a broader component, are a more heterogeneous subclass
than the others. Moreover, the Type IIb SNe, such as SN 1993J in M81
(Filippenko \etal 1993; Nomoto \etal 1993), manifest themselves as
SNe~II at early times, but then experience a spectroscopic
metamorphosis into a SN~Ib at late times (see SN 1987K, Filippenko 1988).

The different subclasses may have quite different progenitor systems.
Thus in this section, we study the photometric and spectroscopic
behavior of SNe 2006my and 2006ov, to get an initial indication of
their nature.

\subsection{Photometry of SN 2006my and SN 2006ov}

We followed SNe 2006my and 2006ov with the 0.76-m Katzman Automatic
Imaging Telescope (KAIT; Li \etal 2000; Filippenko \etal 2001;
Filippenko 2005) at Lick Observatory soon after their discoveries.
KAIT, which conducts the successful Lick Observatory Supernova Search
(LOSS), did not discover these two SNe because of its search strategy
of focusing on near-meridian objects, and the two SNe were 2--3 hours
east of the meridian at the end of the night when discovered by
Itagaki.  As both SNe were discovered at late times, we did not start
an intense multi-color campaign of follow-up observations. Instead,
the two SNe were imaged in unfiltered mode nearly nightly, with
occasional broad-band $BVRI$ photometry.  

An advantage of observing with the unfiltered mode is that we have
good pre-SN unfiltered template images for the SN host galaxies
accumulated during the course of LOSS. These templates enable us to
perform galaxy subtraction, which significantly increases the
photometric accuracy as both SNe (especially SN 2006ov) are heavily
contaminated by host-galaxy emission.  Here we report relative
unfiltered photometry for SNe 2006my and 2006ov, providing the
necessary information on their photometric behavior (e.g., whether
they have a plateau or linear phase). A full treatment of all
available follow-up data for these SNe is beyond the scope of this
paper, and will be discussed elsewhere.

For the unfiltered images of SNe 2006my and 2006ov, we first perform
galaxy subtraction with a customized software package. The SN image is
registered to the pre-SN unfiltered template, their point spread
functions (PSFs) are convolved to the same level, the intensities of
the images are matched, and a final subtraction is performed. This
procedure is essentially that used for processing unfiltered images by
the LOSS SN search pipeline.  We use the IRAF\footnote{IRAF (Image
Reduction and Analysis Facility) is distributed by the National
Optical Astronomy Observatories, which are operated by the Association
of Universities for Research in Astronomy, Inc., under cooperative
agreement with the National Science Foundation.} DAOPHOT package
(Stetson 1987) to perform standard aperture photometry for the SNe and
several bright stars in the field. A relative light curve is then
generated by averaging the differences between the magnitudes of the
SNe and the bright stars.

As the best match to broad-band filters for the KAIT unfiltered data
is the $R$ band (Li \etal 2003), in Figure 3 we compare our relative
unfiltered light curves for SNe 2006my and 2006ov to the $R$-band
light curve of SN 1999em, a well-studied SN~II-P (Leonard \etal 2002b;
Hamuy \etal 2001; Elmhamdi \etal 2003). The relative unfiltered 
light curves of the SNe are
visually shifted by a constant to match the light curve of SN 1999em.
Figure 3 indicates a good match between SN 1999em and SN 2006my: the 
behavior of the decline from the plateau is almost identical for both SNe. 
There is no doubt that SN 2006ov is a SN II-P as well, because it 
shows a prominent plateau phase. However, it has a steeper
decline from the plateau than do SNe 1999em and 2006my. 

Thus, SN 2006my and SN 2006ov behave like SNe~II-P discovered near the 
end of their plateau
phase.  The duration of the plateau phase of SNe~II-P, however, varies
significantly.  From a sample of 13 SNe~II-P studied by Hamuy (2003),
Hendry \etal (2005) derived a mean plateau duration of $131\pm21$
days, with a range of 110 to 170 days.  Consequently, it is unclear
which SN exploded first, SN 2006my or SN 2006ov.  Although SN 2006my
appears to be farther along the plateau than SN 2006ov on the same
date, SN 2006ov could have had a longer plateau phase and thus
exploded earlier.

Our estimate of the age of SNe 2006my and 2006ov at the time of
discovery ($\sim$3 months after explosion) is somewhat older than the
ages suggested by other groups based on spectra: $\sim$1--2 months for
SN 2006my (Stanishev \& Nielsen 2006), and $\sim$1 month for SN
2006ov (Blondin \etal 2006). Next, we will check our own spectra to
provide further age estimates for both SNe, and attempt to constrain
the reddening toward them.

\subsection{Spectroscopy of SN 2006my and SN 2006ov}

We obtained three optical spectra of SNe 2006my and 2006ov with
the Keck 10-m telescopes using the Low Resolution Imaging Spectrometer
(LRIS; Oke et al. 1995), and with the Lick Observatory 3-m Shane
telescope using the Kast double spectrograph (Miller \& Stone
1993). A journal of observations is given in Table 1. All
one-dimensional sky-subtracted spectra were extracted optimally in the
usual manner (e.g., Foley et al. 2003).  For the Kast observations,
flatfields for the red-side CCD were taken at the position of the object to
reduce near-infrared fringing effects.  The spectra were corrected for
atmospheric extinction and telluric bands (Bessell 1999; Matheson et
al. 2000), and then flux-calibrated using standard stars observed at
similar airmass on the same night as the SNe.

  In Figure 4 we show a comparison of the spectra of SNe 2006my and
2006ov to those of other well-studied SNe~II-P at similar epochs. The
spectra of SNe 2004dj and 2003gd are previously unpublished data from
our own spectral database.  All spectra in Figure 4 have been corrected
for the reddening toward the SNe and also for the host-galaxy redshift. For 
SN 2003gd, we adopt $E(B - V) = 0.13$ mag, and an explosion date of 2003
March 17 (Van Dyk, Li, \& Filippenko 2003). For SN 2004dj, we adopt
$E(B - V) = 0.07$ mag, and an explosion date of 2004 June 30 (Zhang
\etal 2005; Vink\'o \etal 2006). We emphasize that both SNe 2003gd and
2004dj were discovered in the middle of the plateau phase, so their
explosion dates have relatively large uncertainties due to the large
scatter in the plateau durations. Only the Galactic reddening of $E(B
- V) = 0.027$ mag and 0.022 mag (Schlegel, Finkbeiner, \& Davis 1998)
were removed from the spectra of SN 2006my and SN 2006ov, respectively.

   Figure 4 demonstrates a striking similarity between the spectra of
SN 2006my and SN 2006ov, as well as their resemblance to the spectra
of SNe 2004dj and 2003gd at 3--4 months after explosion. Other signs
of a relatively old age for these SNe II-P are the strong, relatively 
narrow, Ca~II near-IR triplet (one sees the 8498~\AA\ and 8662~\AA\ 
components, with the 8542~\AA\ component blended within), and
hints of the [Ca~II] $\lambda\lambda$7291, 7324 and [Fe~II]
$\lambda$7155 lines often seen in SNe~II-P during the nebular phase.

   The spectra in Figure 4 also show that with just the corrections
for Galactic reddening, the spectra of SNe 2006my and 2006ov have a
continuum shape similar to that of the other two SNe~II-P. This argues
against the existence of large host-galaxy reddening toward either SN
2006my or SN 2006ov. Blondin \etal (2006), however, suggested that SN
2006ov is reddened (but they did not quote an estimate of the amount
of the host-galaxy reddening). A possible cause of this discrepancy
may be that Blondin \etal compared their spectrum of SN 2006ov to that
of a SN~II-P at a much earlier phase ($\sim$1 month after explosion)
than our estimated age ($\sim$3 months after explosion). As SNe~II-P
become progressively redder at later times, comparing the spectrum of
an older SN~II-P to a younger object would give the incorrect impression
that the older object is reddened. 

  To further investigate the reddening, in Figure 5 we plot a close-up
of the spectral range near the Na~I~D absorption lines of both SN
2006my and SN 2006ov.  It is expected that reddening caused by the
dust in a galaxy would produce noticeable narrow Na~I~D absorption
lines in an object's spectrum, although the quantitative correlation between
the strength of Na~I~D absorption and the amount of reddening is still
quite uncertain. Figure 5 shows no signs of prominent, narrow Na~I~D at 
the rest wavelengths corresponding to either the Milky Way Galaxy or the
hosts (a hint of narrow Na~I~D absorption due to the Milky Way Galaxy
might be present in the spectrum of SN 2006ov).  We conclude that there
is no evidence for significant host-galaxy reddening of either 
SN 2006my and SN 2006ov.

The photometry and spectroscopy of SNe 2006my and 2006ov indicate that
both objects are SNe~II-P discovered near the end of the plateau
phase. We now turn to the analysis of the archival {\it HST} data
available for NGC 4651 and M61, identifying the progenitors of both 
SNe and studying the environments of the progenitors.

\section{The Progenitors of SNe 2006my and 2006ov in Archival {\it HST} Images}

Tables 2 and 3 list the pre-SN archival {\it HST}/WFPC2 data available 
for the host galaxies of SN 2006my (NGC 4651) and SN 2006ov (M61). 
There are also STIS, NICMOS, and ACS/HRC data for M61, but the 
actual site of SN 2006ov was not observed in those cases.  
For the {\it HST}/WFPC2 data, the site of
SN 2006my was imaged in the F555W and F814W filters (Table 2), while the
site of SN 2006ov was imaged in the F450W, F606W, and F814W filters (Table
3). (Hereafter, we will refer to these images by their filter names.)
We downloaded the {\it HST}/WFPC2 data for NGC 4651 and M61 from the {\it
HST} Multimission Archive\footnote{http://archive.stsci.edu/}, and
analyzed them with the STSDAS software package. Cosmic-ray hits in the
CR-split images were removed using the task CRREJ, and the images on
the four individual WFPC2 chips were combined in a mosaic using the
task WMOSAIC.

It is essential to locate with high astrometric precision the SN sites
in the pre-SN {\it HST}/WFPC2 images. For this purpose, we obtained
high-resolution ($0{\farcs}187$ pixel$^{-1}$) images of SNe 2006my and
2006ov with the 3.6-m Canada-France-Hawaii Telescope (CFHT + MegaCam)
in the Sloan $r\arcmin$ band under fair to excellent seeing
conditions (full width at half-maximum intensity about 
$0{\farcs}6$--$1{\farcs}0$). A summary of these CFHT data is 
given in Table 4.

\subsection{Astrometric Solution for NGC 4651 and SN 2006my}

We attempted to match the CFHT $r\arcmin$-band images of SN 2006my to
the mosaic F814W image of NGC 4651. Both sets of CFHT images give 
consistent results, and here we present the results from the 2006 Dec. 24
image due to its superior depth and seeing. We identified 20 stars (or 
compact star clusters) present in both the CFHT and WFPC2 images for 
which we were able to precisely measure the (X,~Y) center positions. 
Then, using the IRAF task GEOMAP, we
performed a geometrical transformation between the two sets of
coordinates, and were able to match them to $\ale 0.98$ WFC pixel
root-mean-square (1$\sigma$ $\ale 0{\farcs}098$). As the mosaic
process involves some uncertainties caused by geometrical distortions
and chip gaps among the four individual WFPC2 chips, we further attempted
to match the CFHT image to the Chip 2 (WFC2) F814W image (where the SN
site is located) before the image was mosaiced.  13 common stars were
identified, with a geometrical transformation uncertainty of 0.45 pixel
(1$\sigma$ $\ale 0{\farcs}045$).

The SN position measured on the CFHT image was then transformed to the
WFPC2 image.  The transformed SN locations based on the geometrical
transformation using the mosaic image and the individual WFC2 image
are consistent with each other within the uncertainties, and we adopt
the SN location based on the WFC2 chip image.  

Figure 6 shows a 5$\arcsec \times 5\arcsec$ close-up of the SN 2006my
environment in the WFPC2 images (the F555W image has the same pointing
as the F814W image).  The white circles in Figure 6 have a radius of
0$\farcs$225 (5 $\times$ 1$\sigma$ error).  Within the 1$\sigma$ error circle,
there is an apparent source in the F814W image, but no apparent point
sources are visible in the F555W image. Figure 6 also shows the images
with a resampled resolution of 0$\farcs$05/pixel (a cubic spline function
is used to interpolate one pixel into $2\times2$ pixels). As can be seen,
resampling the data brings out more details in the undersampled WFPC2
data, and the red source within the error radius of the F814W image is
more easily discerned.  We note that resampling the data has the risk
of smoothing several (extended) sources into a point source, but in
this case the point source is apparent in the original image as
well. 

In both the original and the resampled images, the source appears to
have a slight east-west extension. It is unclear whether this source
is actually a blending of two stars, given the relatively low
signal-to-noise ratio (S/N) for the detection, and given the low spatial
resolution of {\it HST}/WFPC2 at the distance of NGC 4651. The
probability of two RSGs occurring in adjacent pixels on the WFC2
image, however, is very small considering the density of such stars in
the environment of SN 2006my (0.4\%, not including possible physical
clustering of massive stars; see discussion in \S 5).  We
consider this source as a single star, and the likely progenitor of SN
2006my.  The nature of this source is further discussed in \S 4.

In the original non-mosaic {\it HST}/WFPC2 images, the SN site is
located at (X,~Y) = (410.61, 158.81) on Chip 2 (WFC2), with a
1$\sigma$ error circle radius of 0.45 pixel. We note that these
coordinates are in the IRAF system, and need to be adjusted in some
other software packages. For example, the photometry packages Dophot
and HSTphot use a coordinate system that differs from that of IRAF by 
0.5 pixel in both X and Y. In HSTphot, which we use to reduce the
{\it HST}/WFPC2 data in this paper, the SN site is located at
(X,~Y) = (410.11, 158.31) on the WFC2 chip.

\subsection{Astrometric Solution for M61 and SN 2006ov}

We first attempted an astrometric solution for M61 and SN 2006ov
between the CFHT and the pre-SN {\it HST}/WFPC2 F814W images.
25 common stars were used in IRAF/GEOMAP to do a geometrical
transformation with a precision of 1$\sigma$ $\ale 1.10$ WFC pixel
(0$\farcs$11), due to the mediocre seeing of the CFHT image.
Nontheless, after mapping the SN position from the CFHT image onto
the pre-SN F814W image, we identified a possible progenitor within
the 1$\sigma$ error radius, which is subsequently verified  by
the better astrometric solution described below. 

To further improve the astrometric solution, we took images of SN
2006ov with {\it HST}/ACS on 2006 Dec. 12, as part of SNAP program
GO-10877 (PI: W.~Li). The details of the observations are listed in
Table 5. The ACS images were observed with the High Resolution Channel
(HRC) of ACS with a spatial resolution of $0\farcs025$ pixel$^{-1}$
and a field of view of $29'' \times 25''$. We identified 12
common stars in the ACS/HRC F625W image and the Chip 4 (WFC4) pre-SN
WFPC2 F814W image (where the SN site is located), and achieved an
astrometric solution with a 1$\sigma$ error of 0.17 WFC pixel
(0$\farcs$017) (Figure 7).  The F450W image has the same pointing as
the F814W image, but the F606W image does not. An astrometric solution
between the ACS/HRC F625W image and the Chip 4 (WFC4) WFPC2 F606W
image (where the SN site is located) using 9 common stars yields a
precision with a 1$\sigma$ uncertainty of 0.16 WFC pixel
(0$\farcs$016). Using these astrometric solutions, the SN position
measured in the ACS/HRC F625W image is mapped onto the pre-SN WFPC2
images.

Figure 8 shows the 5$\arcsec \times 5\arcsec$ close-up of the SN
2006ov environment in the WFPC2 images. To guide the eye, white
circles with a radius of 20 times the 1$\sigma$ uncertainty of the 
astrometric registration are marked.  Within the very small 1$\sigma$ 
error circle, there is an apparent source in the F814W image, which we
identify as the progenitor of SN 2006ov. A hint of this source can
also be seen in the F606W image, although it appears somewhat
extended. There are also some faint sources near the center of the error
circle in the F450W image, but they appear to be offset from the
progenitor in the F814W image. A bright source is seen to the
northeast of the progenitor in all three bands. It looks stellar
in the F814W image, but is somewhat extended in the F606W and F450W
images, perhaps due to blending with another source. It is also possible
that the source is actually a small star cluster. The presence
of this bright source complicates the analysis of the photometry of
the progenitor of SN 2006ov, as we further discuss in \S 4.

In the pre-SN {\it HST}/WFPC2 F450W and F814W images, the SN site is
located at (X,~Y) = (571.62, 236.22) on Chip 4 (WFC4), with a 1$\sigma$
error circle of 0.17 pixel. In the F606W image, it is at (X,~Y) = (227.44,
267.34) on Chip 4 (WFC4), with a 1$\sigma$ error circle of 0.16
pixel. In the HSTphot coordinate system, these positions are (X,~Y) 
= (571.12, 235.72) for the F450W and F814W images, and (X,~Y) =
(226.94, 266.84) for the F606W image.

\section{The Nature of the Progenitors of SN 2006my and SN 2006ov} 

We have identified the likely progenitors of SNe 2006my and 2006ov in
pre-SN {\it HST}/WFPC2 images. In this section, we study the 
nature of these objects based on their magnitudes and colors.

\subsection{Photometry of the Supernova Progenitors}

We used the software package HSTphot (Dolphin 2000a, 2000b) to conduct
photometry of the {\it HST}/WFPC2 images of SNe 2006my and 2006ov.
HSTphot automatically accounts for WFPC2 PSF variations and
charge-transfer effects across the chips, zeropoints, aperture
corrections, etc.  There are many option flags to run HSTphot. For our
reduction, we chose to include Option 2 (turn on local sky
determination) as recommended by the HSTphot manual for images of
galaxies well beyond the Local Group, and Option 8 (turn off aperture
corrections) as there are no good aperture correction stars in our
images. HSTphot then uses the default aperture corrections for the
filters, which are probably accurate in general to 0.02 mag. We also
used an independent detection threshold of 2.5$\sigma$ (minimum S/N
for a given image or filter for star detection), and a total detection
threshold of 3.0$\sigma$ (minimum total S/N for a star to be kept in
the final output).  All photometry was performed on the coadded images
in each filter.

Table 6 lists the HSTphot photometry for the progenitors of SNe
2006my and 2006ov. Columns 2 and 3 list the SN location as predicted
by the astrometric solutions (\S 3). Columns 4 and 5 list the
coordinates of the detected sources near the SN location in
HSTphot. From the magnitudes of all 3$\sigma$ detections in the 
images, we also empirically determined the 
limiting magnitude near the SN sites in the images, and listed
them in Column 9 (Lmag) of Table 6. These limiting magnitudes are 
different ($\sim$0.5 to 1.0 mag shallower) from the values calculated
with the WFPC2 ETC on the {\it HST} 
website\footnote{http://www.stsci.edu/hst/wfpc2/software/wfpc2-etc.html}, 
as the SN sites are located on a bright stellar background.

For SN 2006my in NGC 4651, HSTphot detected a 5.6$\sigma$
source close to ($\ale 0.43$ pixel), and within the 1$\sigma$ error of,
the predicted SN location in the F814W image, with a flight system
magnitude of F814W = $24.47 \pm 0.20$.  For the F555W image, HSTphot
failed to detect any source within the 1$\sigma$ error radius (the
closest detection is an extended source near the $\sim 2\sigma$ error
radius).  Inspection of Figure 5 indicates that there is no apparent
point source within the 1$\sigma$ error radius. As HSTphot has an
option to accept pre-determined star lists, we enforced HSTphot to
make a measurement at the position of the progenitor as measured by
HSTphot in the F814W image; this yielded a 2.2$\sigma$ detection with a
flight system magnitude of F555W = $26.66 \pm 0.55$. As this F555W
magnitude is fainter than the 3$\sigma$ limit (F555W $<$ 26.5 mag)
of the image, we consider the progenitor to be not detected in the 
F555W image, and use the 3$\sigma$ limiting magnitude ($V < 26.5$ mag) 
as the upper limit for the $V$-band flux of the progenitor. 

For SN 2006ov in M61, HSTphot failed to detect any sources within the
1$\sigma$ error radius of the SN location in the F814W image. Lowering
the total detection threshold to 2$\sigma$ did not help.  Inspection
of the residual image generated by HSTphot after subtracting the PSFs
of all positive detections, however, reveals an apparent source close
to the known SN location.  Contamination by the bright source to
the northeast of the SN location is a possible explanation for the
failure by HSTphot to make a detection of the central source in the
error circle. 

Since we have an accurate (1$\sigma$ = 0.17 pixel) position for the 
putative progenitor of SN 2006ov as determined from the ACS/HRC to WFPC2
astrometric registration, we forced HSTphot to make a measurement at
the progenitor position.  This procedure yielded a 6.1$\sigma$
detection with F814W = $23.19 \pm 0.18$ mag, and a residual image with
all sources within the 1$\sigma$ error radius cleanly removed. We
further forced HSTphot to make a measurement at the progenitor
position in the F450W and F606W images.  We measured a 2.2$\sigma$
source with F606W = $24.07 \pm 0.50$, and a 6.1$\sigma$ source with
F450W = $23.51 \pm 0.18$ mag. Although the detection in the F606W image
is of low S/N (2.2$\sigma$) due to possible blending
from other sources, the measured magnitude (F606W = $24.07 \pm 0.50$ mag)
is brighter than the 3$\sigma$ limiting magnitude (F606W $<$ 25.3 mag),
and inspection of Figure 8 suggests a hint of the progenitor
in the F606W image, so we kept the F606W detection as it is. 
The 6.1$\sigma$ detection in the F450W image is likely caused by 
another source; as previously noted, it appears to be offset from the
progenitor in the F814W image. Moreover, if it were indeed the progenitor,
then the spectral energy distribution would be unrealistic, with 
a red $V-I$ color and a blue $B-V$ color.
We assume that, because of the spatial offset, this blue source does 
not substantially affect our $V$ and $I$ measurements of the putative 
progenitor, though there might be some contamination.

The flight-system magnitudes were then transformed to the standard broad-band
$BVRI$ system following the prescriptions by Holtzman \etal (1995) and 
Dolphin (2000b), and are listed in Column 8 (Mag2) of Table 6. 

\subsection{Properties of the Supernova Progenitors}

We can estimate the masses of the progenitor stars by comparing the
intrinsic colors and absolute magnitudes of the objects with stellar
evolution tracks of massive stars having different zero-age
main-sequence masses ($M_{\rm ZAMS}$).  In \S 4.1 we measured the
apparent magnitudes of the possible progenitors (see Table 6 for a
summary). To estimate the absolute magnitudes of the progenitors, we
need to know the distances to NGC 4651 and M61.

For NGC 4651, Solanes \etal (2002) collected 7 distance measurements
derived by different groups using the Tully-Fisher (T-F) method, and
reported an average distance modulus of $\mu = 31.74 \pm 0.25$ mag.
Terry, Paturel, \& Ekholm (2002) reported a ``sosie galaxy" T-F
distance of $\mu = 31.86 \pm 0.17$ mag. We adopt $\mu = 31.8 \pm 0.3$
mag as a possible distance to NGC 4651.

For M61, the T-F distance published by Tully (1988) is $\mu = 30.91$
mag, while Sch\"oniger \& Sofue (1997) reported a CO and H~I T-F
distance of $\mu = 30.12 \pm 0.10$ mag. We adopt $\mu = 30.5 \pm 0.4$
mag as a possible distance to M61.

Both NGC 4651 and M61 may be members of the Virgo cluster (VC), one of
the nearest rich clusters in the northern hemisphere with over 1300
member galaxies, so we also attempted to use the measured distance to
the VC as the distance of both NGC 4651 and M61.  The VC, however, is
quite extended in size and has complicated three-dimensional
structure. The two main components are the somewhat more nearby,
northern, M87 subcluster (dominated by early-type galaxies), and the
somewhat more distant, southern, M49 galaxy concentration (rich in
spiral galaxies).  Currently, there is a total of 8 VC spiral galaxies
with Cepheid distances (NGC 4496A, NGC 4536, NGC 4548, NGC 4321, NGC
4535, NGC 4639, NGC 4527, NGC 4414; see the summary by Freedman \etal
2001), and their average distance is $\mu = 31.03 \pm 0.12$ mag.  A
recent $B$-band T-F distance measurement to 51 VC spiral galaxies
yields $\mu = 31.28 \pm 0.14$ mag (Fouque \etal 2001). Two recent
surface brightness fluctuation (SBF) measurements to the elliptical
galaxies in the VC yield $\mu = 31.09 \pm 0.15$ mag (Tonry \etal 2001;
Jerjen \etal 2004). The distance to the VC galaxy IC~3338 using the
tip of the red giant branch (TRGB) method gives $\mu = 30.98 \pm 0.19$
mag (Harris \etal 1998). Here we adopt $\mu = 31.1 \pm 0.2$ mag as the
average distance to VC. To account for the line-of-sight diameter of the
VC, we further adopt $\mu = 31.1 \pm 0.5$ mag as another possible
distance to both NGC 4651 and M61.

We also note that NGC 4651 is $\sim$4 degrees north of the M87
subcluster, while M61 is $\sim$8 degrees south of the M87 subcluster
(and thus possibly a member of the M49 subcluster).  At face value,
our adopted individual distances of NGC 4651 and M61 ($\mu =
31.8 \pm 0.3$ mag and $30.5\pm0.4$ mag, respectively) contradict the
depicted structure of the VC: NGC 4651 is in the more nearby northern
region but it has a larger adopted distance than the average VC, M61
is in the more distance southern subcluster but it has a smaller
adopted distance than the average VC. As discussed by Pilyugin,
V\'ilchez, \& Contini (2004), however, it seems impossible to check
the distance of individual VC galaxies by comparing their value with
the ``group" distance, largely because of the complicated
three-dimensional structure of the cluster.

The evolutionary tracks of massive stars on the color-magnitude 
diagrams (CMDs) are significantly
affected by the adopted metallicity, so we attempt to constrain the
metallicity of the sites of SNe 2006my and 2006ov by using information
from the literature. From the CFHT images, we measured that SN 2006my
is 27$\farcs$4 west and 22$\farcs$1 south of the nucleus of NGC
4651, and that SN 2006ov is 6$\farcs$1 east and 51$\farcs$1 north
of the nucleus of M61. With these offsets, the galaxy position angle
and the inclination angle from
LEDA\footnote{http://leda.univ-lyon1.fr/}, and the published
metallicity and its radial gradient in NGC 4651 and M61 by Polyugin
\etal (2004), we derived the relative oxygen abundance log(O/H) + 12 =
$8.51 \pm 0.06$ dex and $8.64 \pm 0.10$ dex for the sites of SN 2006my and
SN 2006ov, respectively. These metalicities are lower than the solar
value (8.8 dex; Grevesse \& Sauval 1998), so we adopt a subsolar
metallicity of $Z = 0.008$ for the environment of both SNe.

Figure 9 shows the $(V-I)^0$ vs. $M_I^0$ CMD for the progenitor of SN
2006my in NGC 4651, compared with model stellar evolution tracks for a
range of masses from Lejeune \& Schaerer (2001), assuming enhanced
mass loss for the most massive stars and a metallicity of $Z =
0.008$. We have corrected the magnitudes of the progenitor with the
Galactic reddening only [$E(B-V) = 0.027$ mag], as discussed in 
\S 2.2. The uncertainties in
the photometry and the distance are added in quadrature to produce the
final uncertainty for the absolute magnitude. As a result of only an upper
limit to the brightness of the progenitor in the $V$ band, we have
a lower limit to the $(V-I)^0$  color, as illustrated by the arrows  
in Figure 9.  The filled square represents the data with an adopted 
distance modulus of $\mu = 31.8 \pm 0.3$ mag. 
From the location of the progenitor on the CMD,
$M_{\rm ZAMS} = 11$--15 $M_\odot$ is estimated. The filled circle
represents the data when the average VC distance ($\mu = 31.1 \pm 0.5$
mag) is used, which suggests $M_{\rm ZAMS} = 7$--12 $M_\odot$.

Figure 10 shows the $(V-I)^0$ vs. $M_I^0$ CMD for the progenitor of SN
2006ov in M61, after correcting the magnitudes of the progenitor 
with the Galactic reddening of $E(B - V)$ = 0.022 mag.
Adopting $\mu = 30.5 \pm 0.4$ mag, the $M_{\rm ZAMS}$
estimate for the progenitor is 13--17 $M_\odot$. With the average VC
distance ($\mu = 31.1 \pm 0.5$ mag), we find $M_{\rm ZAMS} = 15$--19
$M_\odot$.  Figure 11 shows the $(V-I)^0$ vs. $M_V^0$ CMD for the same
object. The $M_{\rm ZAMS}$ estimate for the progenitor is 12--17
$M_\odot$ and 14--20 $M_\odot$ for our two choices of the distance,
respectively.

It is clear that the adopted distances to the galaxies have a relatively
large impact on the derived progenitor masses. As a byproduct of the 
CMD study, we provide an independent constraint on the distances to NGC 4651
and NGC 4303 from the global stellar photometry on the {\it HST} images. 
For SN 2006my, we measured the photometry of all the stars (with 
S/N $\age$ 3) on the four WFPC2 chips on the F555W and F814W images,
and plotted their magnitudes on the $(V-I)^0$ vs. $M_I^0$ CMD 
with an adopted distance to NGC 4651. The isochrones of the stellar
evolutionary tracks from Lejeune \& Schaerer (2001) are then plotted
on the CMD to see whether they are a good fit to the data. Although
this method is plagued by the need to adopt a global metallicity for
all the stars (while in reality they should come from a range of 
metallicities from different regions of NGC 4651), possible contamination
of compact stellar clusters in the photometry, and no reddening corrections,
it nevertheless suggests that $\mu = 31.1 \pm 0.5$ mag (the average VC
distance) provides a better fit to the data than the average T-F
distances ( $\mu = 31.8 \pm 0.3$).  For SN 2006ov, the F450W and 
F814W images were used to derive a $(B-I)^0$ vs. $M_I^0$ CMD,
and the data favor the average T-F distance ($\mu = 30.5 \pm 0.4$ mag)
over the average VC distance ($\mu = 31.1 \pm 0.5$ mag). The global metallicity
was adopted as solar for both galaxies, even though the SN sites are subsolar. 

Owing to the relatively large uncertainties in the distance estimates,
our conservative approach is to adopt the progenitor masses as derived
from the distances favored by the CMD study ($\mu = 31.1 \pm 0.5$ mag for
SN 2006my, and $\mu = 30.5 \pm 0.4$ mag for SN 2006ov), but allow the
errorbars to cover the range of progenitor masses from the other distance
estimate.  This gives $M_{\rm ZAMS}$ = $10^{+5}_{-3}$ $M_\odot$ for the
progenitor of SN 2006my in NGC 4651, and $15^{+5}_{-3}$ $M_\odot$ for the
progenitor of SN 2006ov in M61.  Both progenitors have a red color,
albeit with a large uncertainty. Their locations on the CMDs suggest
that the progenitor of SN 2006my is a RSG, while the progenitor of
SN 2006ov is consistent with either a RSG or a YSG.

\section{Discussion}

\subsection{Progenitors of SN 2006my and SN 2006ov}

We have located the sites of SNe 2006my and 2006ov to high precision
on pre-SN {\it HST}/WFPC2 images, and identified the likely progenitor
stars of both SNe. As can be seen in Figures 1 and 2, both SNe
occurred on a spiral arm of their host galaxies, and their immediate
environments are heavily contaminated by host-galaxy emission
(more so for SN 2006ov than SN 2006my). Even at {\it HST}/WFPC2
resolution (Figures 5 and 8), there are many sources in the SN
environments.  It is thus possible that the objects we identified in
the error circles are just RSGs that happened to be close to the SN
explosion, but are completely unrelated.  

The confidence of the progenitor identifications can be strengthened if
we put stringent limits on the non-detection of the progenitors.
Using the limiting magnitudes in Table 5, we place the following mass
limits if the progenitors of SNe 2006my and 2006ov were not detected
in the {\it HST}/WFPC2 F814W images: $M_{\rm ZAMS} \ale$ 11 $M_\odot$ or
8 $M_\odot$ for the SN 2006my progenitor, when $\mu = 31.8 \pm 0.3$
mag and $\mu = 31.1 \pm 0.5$ mag are adopted for NGC 4651,
respectively; $M_{\rm ZAMS} \ale$ 9 $M_\odot$ or 13 $M_\odot$ for the SN
2006ov progenitor, when $\mu = 30.5 \pm 0.4$ mag and $\mu =
31.1 \pm 0.5$ mag are adopted for M61, respectively. These mass limits
are not very restrictive, unfortunately, as several of the previously
identified progenitors for SNe~II-P have masses in the range 8--13
$M_\odot$ (Van Dyk \etal 2003a; Smartt \etal 2004; Maund \etal 2005a;
Li et al. 2006; Hendry \etal 2006). Hence, there exists the possibility
that the progenitors of SN 2006my and/or SN 2006ov were not detected.

The probability of a chance coincidence (that is, the objects we
identified are unrelated RSGs that happened to be within the error
circles) can be further assessed by studying the density of RSGs in
the SN environments. Within a 1$\farcs$0 radius (approximately 80 pc
at our adopted average VC distance), we identified 6 possible RSGs (8
possible RSGs) in
the neighborhood of SN 2006my (SN 2006ov) by noting their absolute
magnitudes and colors on the CMDs.  The probability of a RSG within
the 0$\farcs$045 (0$\farcs$02) error radius is thus 1.2\%
(0.2\%). The small probability (0.2\%) of a chance coincidence for the
progenitor of SN 2006ov solidifies our identification. While the
probability of a chance coincidence for the progenitor of SN 2006my is
also small (1.2\%), the confidence of our progenitor identification
can be further increased if we can significantly reduce the astrometric
error radius.  

In \S 3.1 we also mentioned a possible blending issue for the
progenitor of SN 2006my. Using the RSG density in the environment of
SN 2006my, the chance of two RSGs occurring in adjacent WFC pixels
(0$\farcs$1) is only 0.4\%, assuming that massive stars don't tend
to occur in clusters. (The probability of blending would increase 
somewhat if the tendency of massive stars to occur in associations
were included, but it would still likely be relatively low.)

The two main reasons for the large mass range of our identified
progenitors are (a) the uncertain distances to the galaxies, and (b)
the large photometric uncertainty due to the low S/N of the
detections.  When more accurate distances for the galaxies (e.g.,
Cepheid distances) become available, the masses of the progenitors will
be further constrained. The second problem highlights the difficulty
in directly identifying progenitors of SNe: they are faint beyond the
Local Group, and the pre-SN {\it HST} archival images often do not
have the optimal combination of filters and depth because most of them
were obtained by various observers for different projects.  A
dedicated {\it HST} program to image the most nearby
galaxies would give observers a chance to identify the progenitors of
most core-collapse SNe in these galaxies, further advancing our
understanding of the death of massive stars.

\subsection{The Nature of Core-Collapse Supernova Progenitors}

Our identified progenitors of SN 2006my and SN 2006ov increase the
number of directly detected progenitors for genuine SNe from 7 to
9. The first half of Table 6 lists the inferred masses of these
progenitors.  The progenitors of several probable super-outbursts of
luminous blue variables (LBVs) misclassified as SNe, such as SN 1961V
(Zwicky 1964) and SN 1997bs (Van Dyk \etal 1999b, 2000), 
have been identified but are not listed. 

The second half of Table 6 gives the mass limits for the progenitors
of several core-collapse SNe. For these SNe, the progenitors are not
directly identified, but a mass limit is derived based on the limiting
magnitudes of pre-SN images. In particular, SN 2004dj occurred in a
compact star cluster (CSC), and the mass of the progenitor is
estimated from the properties (luminosity and color) of the CSC
(Ma\'iz-Apell\'aniz \etal 2004; Wang \etal 2005). SN 2005gl is
associated with a very luminous ($M_V = -10.3 \pm 0.2$ mag) source
(Gal-Yam \etal 2007), which could be a very massive single star (e.g.,
an LBV) or a CSC. There is also some uncertainty regarding the exact
type of the SN: while a very early-time spectrum showed narrow hydrogen
lines typical of SNe~IIn, its later spectral evolution is more
typical of normal SNe~II, perhaps a SN~II-L.

It is clear from Table 6 that observers have had the most success
identifying progenitors for SNe~II-P (7 out of 9). The main reason for
this is perhaps the relative frequency of different types of SNe.  A
preliminary analysis indicates that out of the 68 core-collapse SNe
discovered by LOSS within 30 Mpc in the past 9 years, 46 are SNe~II
(most of which are SNe~II-P), 18 are SNe~Ib/c, 1 is a SN~IIb, and 3 are
SNe~IIn.  SNe~II-P are thus by far the most common type of
core-collapse SN. Still, it is a bit surprising that no direct
progenitor has been detected for a SN~Ib/c, given that they are
roughly 40\% as frequent as SNe~II from the LOSS statistics. One
factor is perhaps the low luminosity of the possible progenitors for
SNe~Ib/c, which also limits the progenitors to be either relatively 
low-mass stripped stars in binary systems, or single Wolf-Rayet 
stars.

The inferred masses for the progenitors of SNe 2006my and 2006ov are 
consistent with the trend that SNe~II-P come from low-mass (8--20 $M_\odot$)
RSGs, as previously suggested (Li \etal 2006; Hendry \etal 2006).
In total, there are now 7 directly identified progenitors for 
SNe~II-P, all with $M_{\rm ZAMS} \ale 20 M_\odot$. The statistics are
growing, and suggest that perhaps all SNe~II-P come from low-mass RSGs. 
If this is true,  stars more massive than $\sim20~M_\odot$
may have a different evolutionary path, and explode as other types of 
SNe such as SNe~II-L or SNe~IIn. The possible association of the 
SN~IIn/II-L 2005gl with a very massive single star (Gal-Yam \etal 2007)
is in agreement with this speculation. In the future, when a
very massive star explodes as a SN in a nearby galaxy (a rare 
event), observers may have a chance to study both the massive star and
the SN in detail. Until then, the fate of the very massive stars
($M_{\rm ZAMS} > 20 M_\odot$) still needs to be observationally verified. 

\section{Conclusions}

\begin{enumerate}

\item{Both SN 2006my in NGC 4651 and SN 2006ov in M61 are SNe~II-P
discovered 3--4 months after explosion. Their spectra show no 
evidence of strong host-galaxy reddening.} 

\item{We obtained high-quality ground-based CFHT images for both SNe,
and were able to locate the SN sites to high precision on the pre-SN
{\it HST}/WFPC2 images (astrometric uncertainty 1$\sigma$ = 0$\farcs$045
and 0$\farcs$11 for SNe 2006my and 2006ov, respectively). We further
improved the astrometric solution to 1$\sigma$ = 0$\farcs$02 for SN 
2006ov with {\it HST}/ACS images of the SN. Within 
each error circle, we identified a likely progenitor of each SN.}

\item{Photometric analysis suggests that the progenitor of
SN 2006my has $M_{\rm ZAMS} = 10^{+5}_{-3}~M_\odot$, and the progenitor
for SN 2006ov has $M_{\rm ZAMS} = 15^{+5}_{-3}$ $M_\odot$. There is a small
probability (1.2\% and 0.2\%, respectively) that the object we identified
as the progenitor is caused by chance coincidence.} 

\item{The inferred masses for the possible progenitors of SNe 2006my
and 2006ov are consistent with the trend that the most common core-collapse
SNe~II-P arise from stars with $M_{\rm ZAMS}$ in the range 8--20 $M_\odot$.}

\end{enumerate}

\bigskip

\acknowledgments

   We thank Louis Desroches, Mohan Ganeshalingam, Jennifer Hoffman,
Douglas Leonard, Frank Serduke, Jeffery Silverman, and Diane Wong for
obtaining and reducing the optical spectra of SNe used in this paper
with the 3-m Shane reflector at Lick
Observatory.  The work of A.V.F.'s group at U. C. Berkeley is
supported by National Science Foundation grant AST-0607485, and by the
TABASGO Foundation.  Additional funding is provided by NASA through
grants AR-10690, AR-10952, and GO-10877 from the Space Telescope
Science Institute, which is operated by the Association of
Universities for Research in Astronomy, Inc., under NASA contract
NAS~5-26555. KAIT was made possible by generous donations from Sun
Microsystems, Inc., the Hewlett-Packard Company, AutoScope
Corporation, Lick Observatory, the National Science Foundation, the
University of California, and the Sylvia \& Jim Katzman Foundation.

\newpage

\renewcommand{\baselinestretch}{1.0}

\newpage

\begin{figure*}
{\plotfiddle{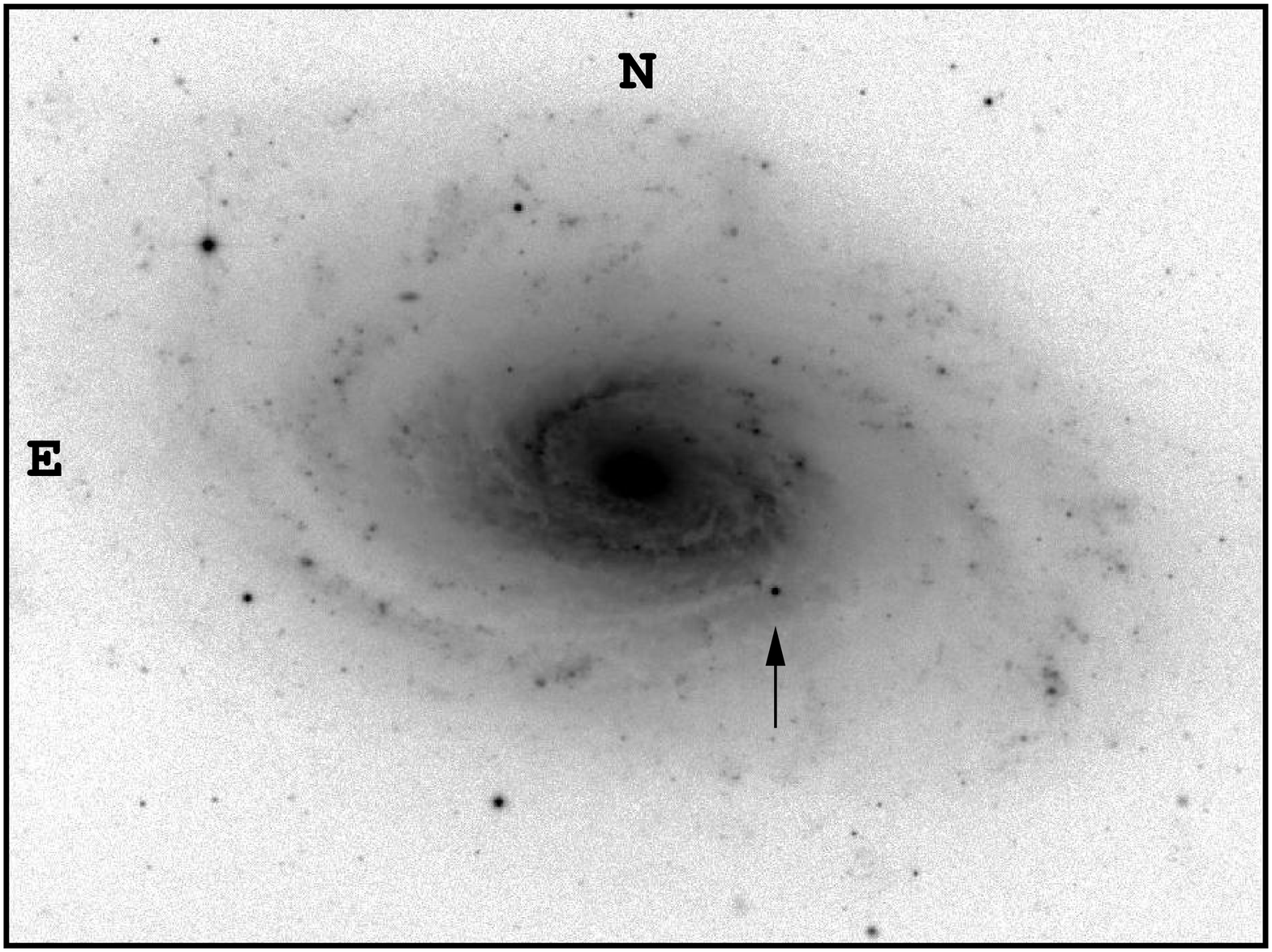}{6.4in}{0}{50}{50}{-40}{100}}
\caption[] {
A $4\arcmin \times 3\arcmin$ section of the CFHT $r\arcmin$-band MegaCam image 
of SN 2006my, taken under 0$\farcs$6 seeing on 2006 Dec. 24. SN 2006my is 
marked with an arrow. 
}
\label{1}
\end{figure*}

\begin{figure*}
{\plotfiddle{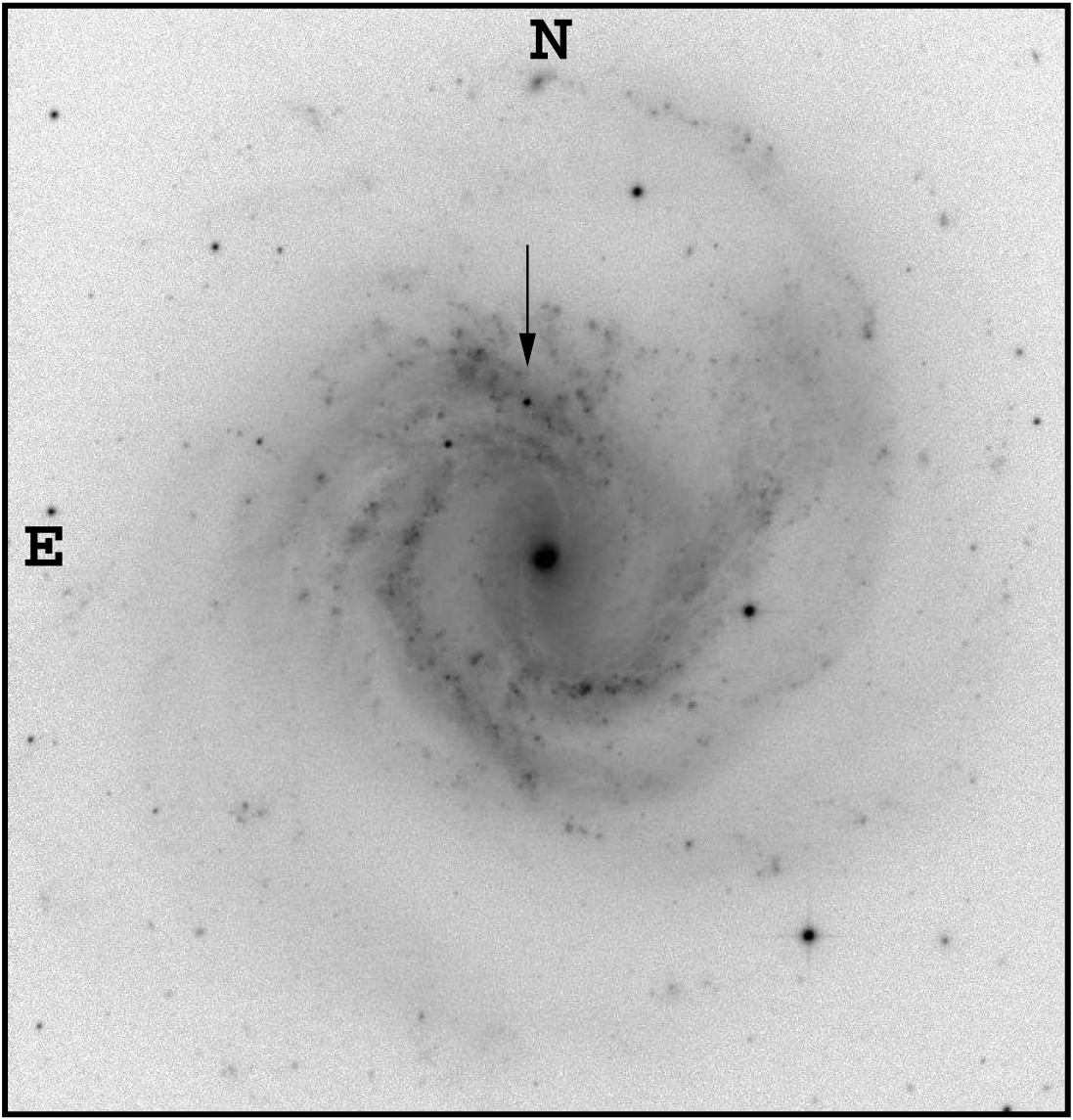}{6.4in}{0}{50}{50}{-40}{100}}
\caption[] {
A $6\arcmin \times 6\arcmin$ section of the CFHT $r\arcmin$-band MegaCam image
of SN 2006ov, taken under 1$\farcs$0 seeing on 2006 Nov. 27. 
SN 2006ov is marked with an arrow.
}
\label{2}
\end{figure*}

\begin{figure*}
{\plotfiddle{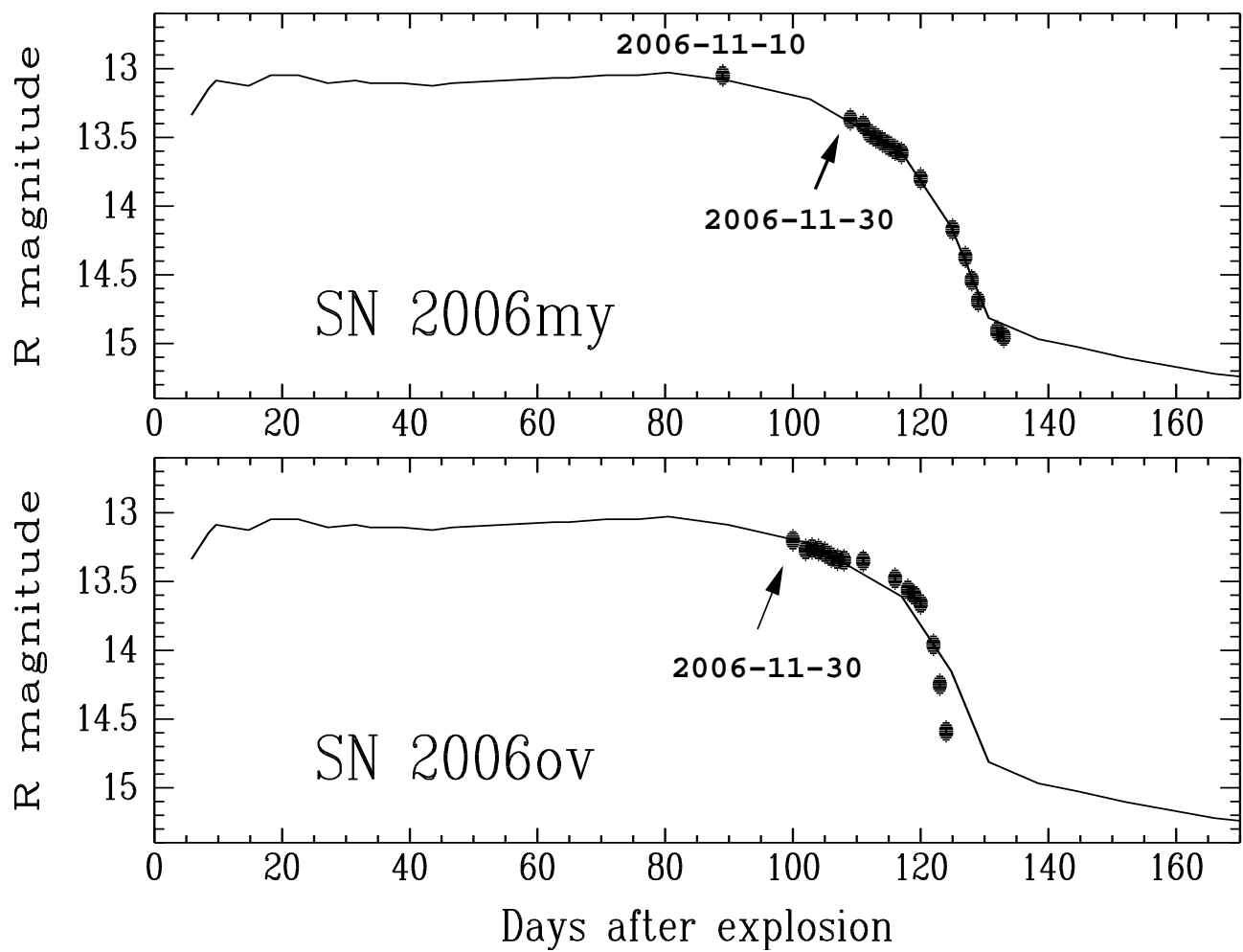}{6.4in}{0}{70}{70}{-40}{100}}
\caption[] {
Comparison of the light curves of SNe 2006my and 2006ov to that of SN 1999em
(Hamuy \etal 2001; Leonard \etal 2002b). The photometry of SNe 2006my and 
2006ov was performed on unfiltered images, and the data were visually
shifted by a constant to match SN 1999em (whose light curve is plotted as a solid line). Dates
of some observations
are marked. Both SNe show characteristics of SNe~II-P discovered
near the end of the plateau phase.
}
\label{3}
\end{figure*}

\begin{figure*}
{\plotfiddle{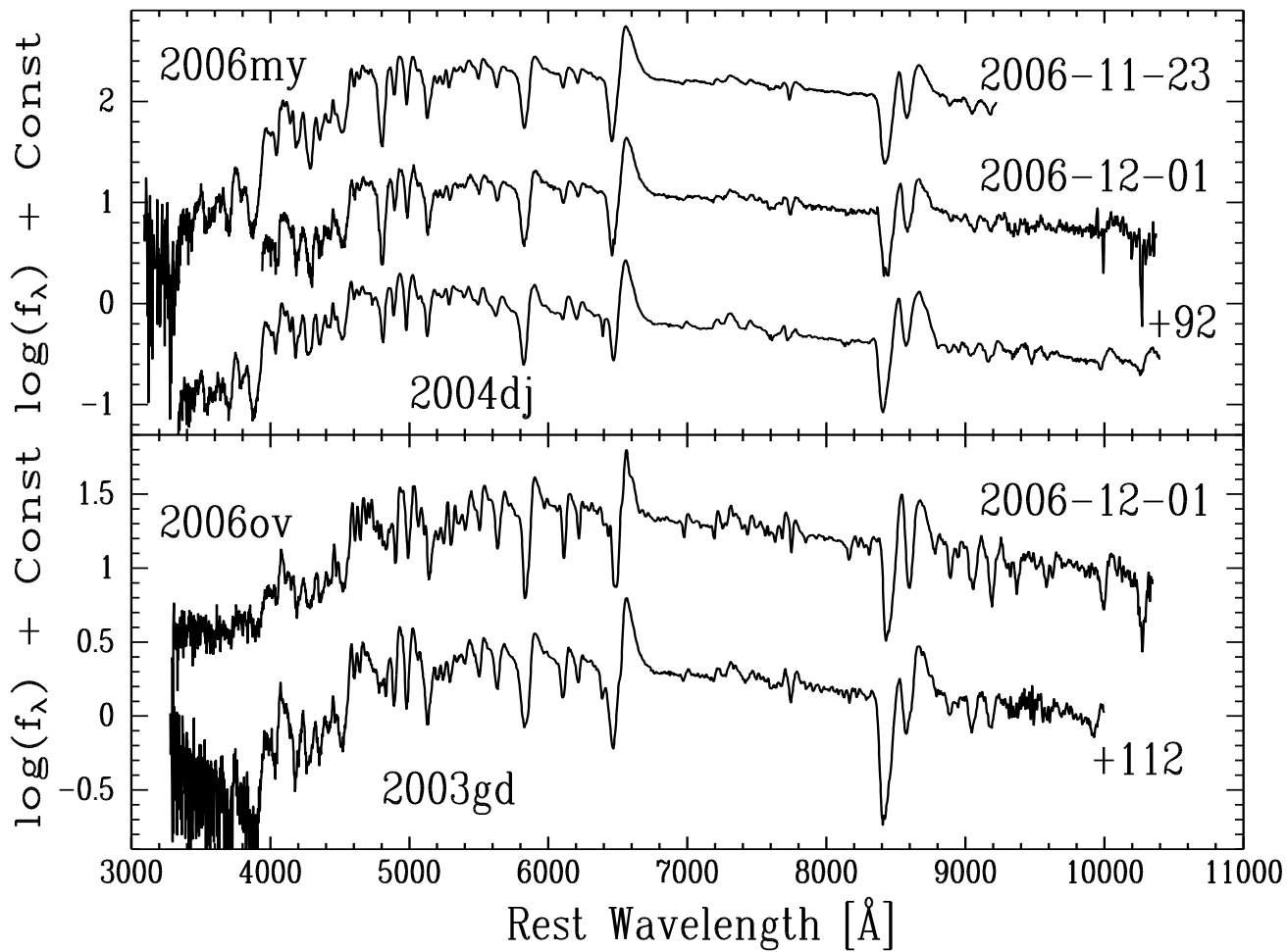}{6.4in}{-90}{70}{70}{-40}{500}}
\caption[] {
Comparison of the spectra of SNe 2006my and 2006ov to those of other 
well-studied SNe~II-P at similar epochs. The spectra have been 
corrected for reddening and host-galaxy redshift (see text for details). 
}
\label{4}
\end{figure*}

\newpage

\begin{figure*}
{\plotfiddle{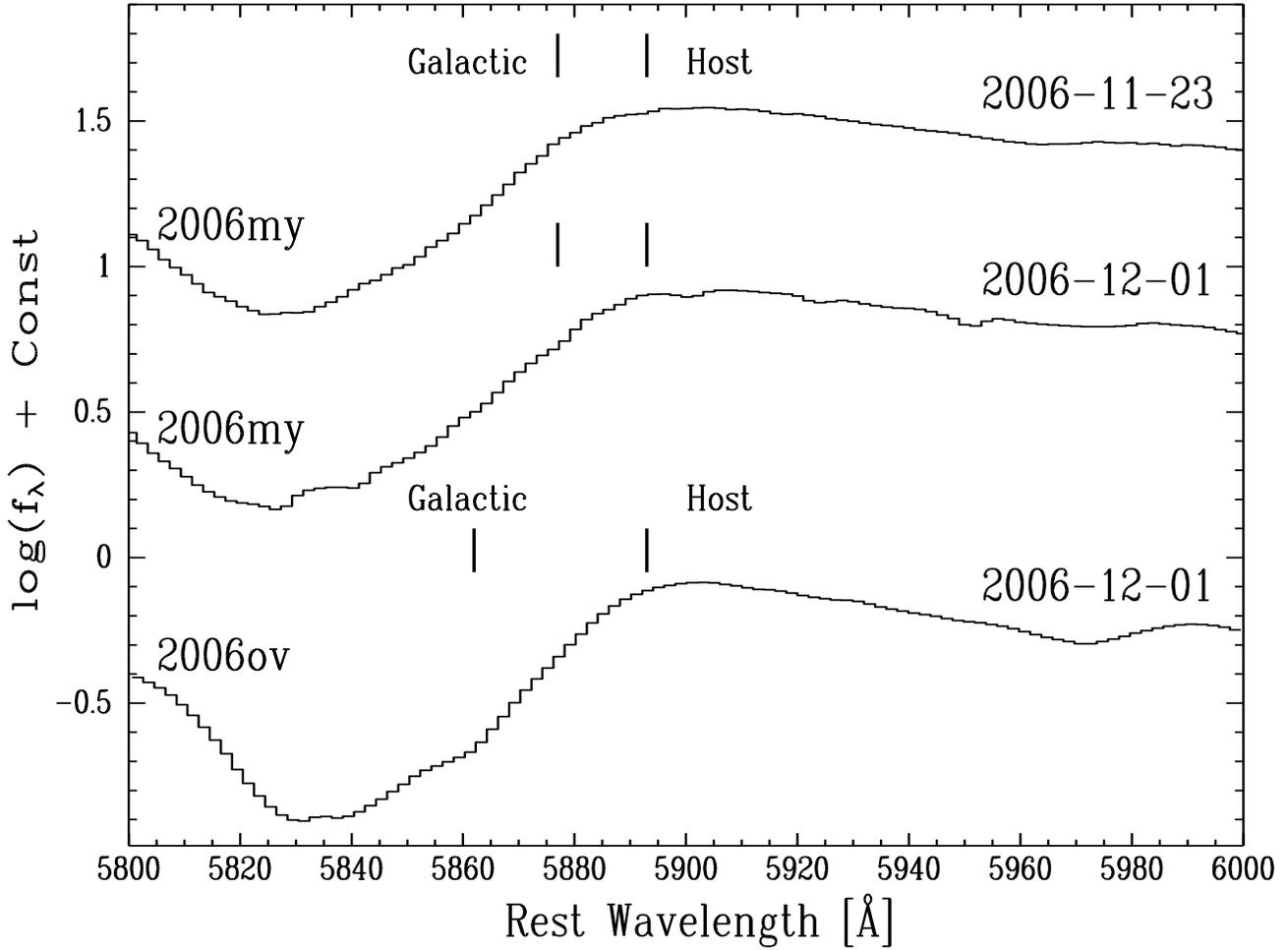}{6.4in}{-90}{70}{70}{-40}{500}}
\caption[] {
A close-up of the spectra of SNe 2006my and 2006ov. The expected 
positions of the interstellar Na~I~D absorption lines at the 
rest wavelength (Galactic lines) and at the redshift of the SN 
(host-galaxy lines) are marked. No apparent Na~I~D absorption 
lines due to the host galaxies are observed. 
}
\label{5}
\end{figure*}

\begin{figure*}
{\plotfiddle{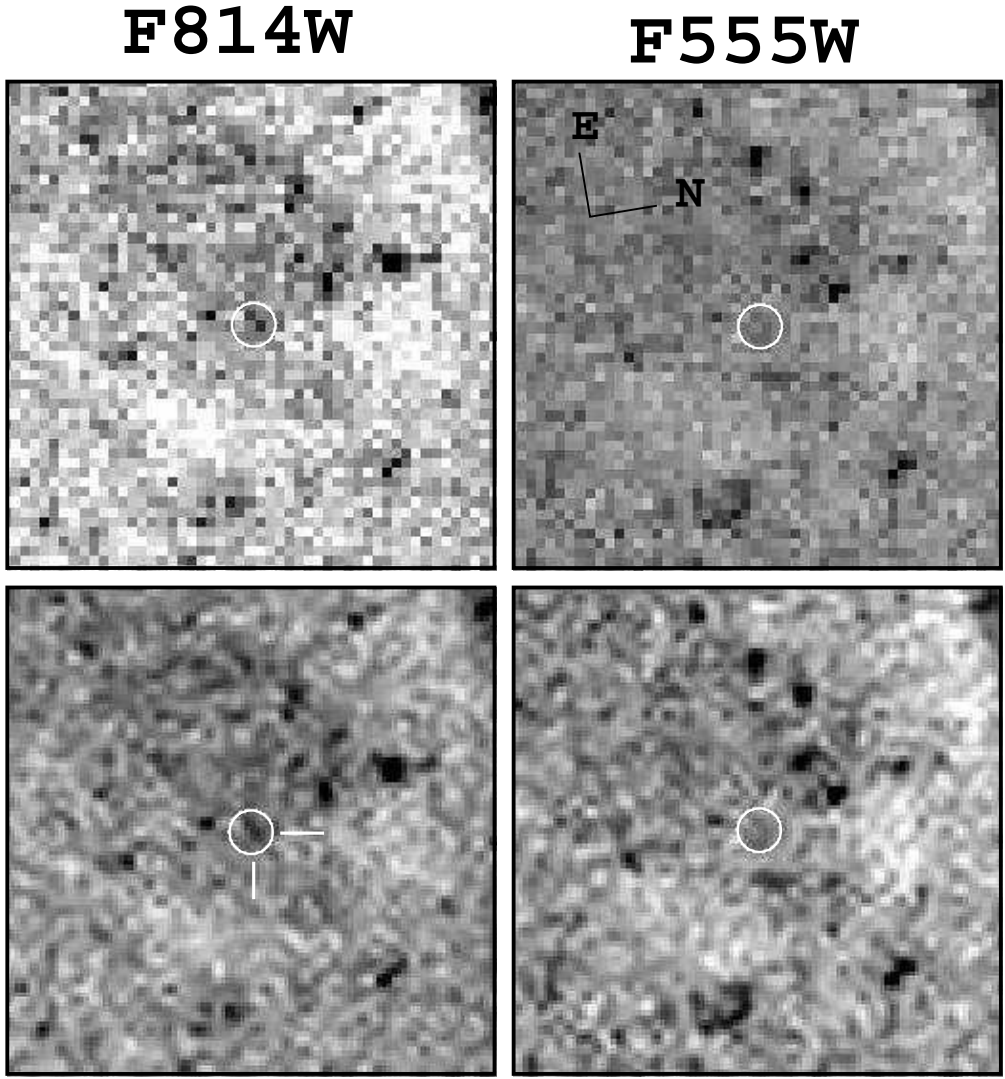}{6.4in}{0}{80}{80}{0}{60}}
\caption[] {
A $5\arcsec \times 5\arcsec$ close-up of the SN 2006my environment in the
$HST$/WFPC2 F814W and F555W images. The white circles mark 5 times the
$1\sigma$ uncertainty of the astrometric registration. The upper panel 
shows the images in the original WFPC2 resolution (0$\farcs$1/pixel), 
while the lower panel resamples the data to a resolution of 
0$\farcs$05/pixel to bring out more details. The candidate progenitor
is marked by a cross hair in the resampled F814W image. Photometry of the 
stars was performed on the original images. 
}
\label{6}
\end{figure*}

\begin{figure*}
{\plotfiddle{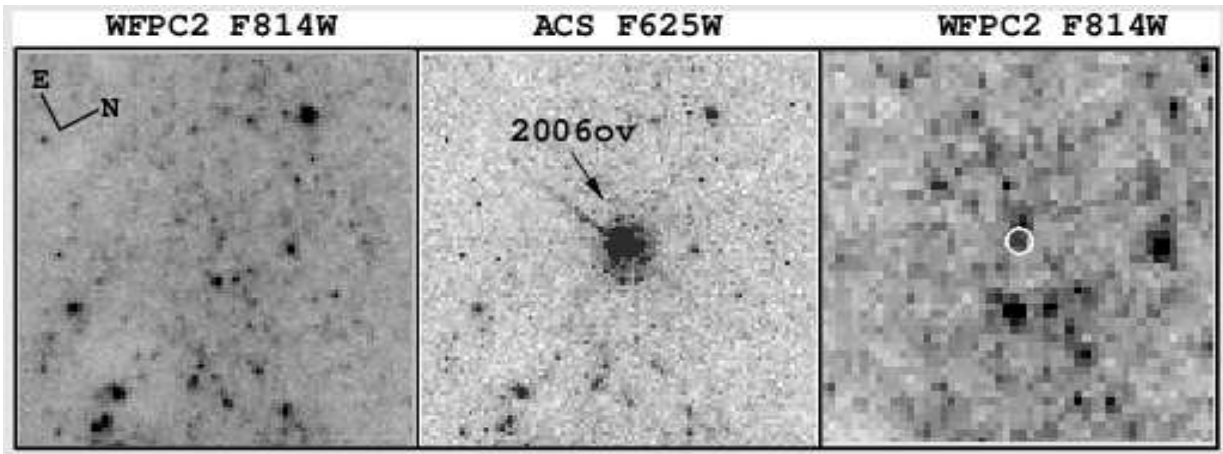}{6.4in}{0}{100}{85}{-20}{70}}
\caption[] {{\it Left:} A 10$\arcsec \times 10\arcsec$ section of the
pre-SN $HST$/WFPC2 F814W image of the SN 2006ov region. 
{\it Center:} A 10$\arcsec \times 10\arcsec$
section of the $HST$/ACS F625W image with SN 2006ov, carefully registered
to the image shown in the left panel. 
{\it Right:} A 5$\arcsec \times 5\arcsec$ close-up of the SN 2006ov
environment. The white circle marks 10 times the $1\sigma$ uncertainty
of the astrometric registration. An apparent stellar object is seen in
the center of the circle, although it is contaminated by a nearby
bright source. 
}
\label{7}
\end{figure*}

\begin{figure*}
{\plotfiddle{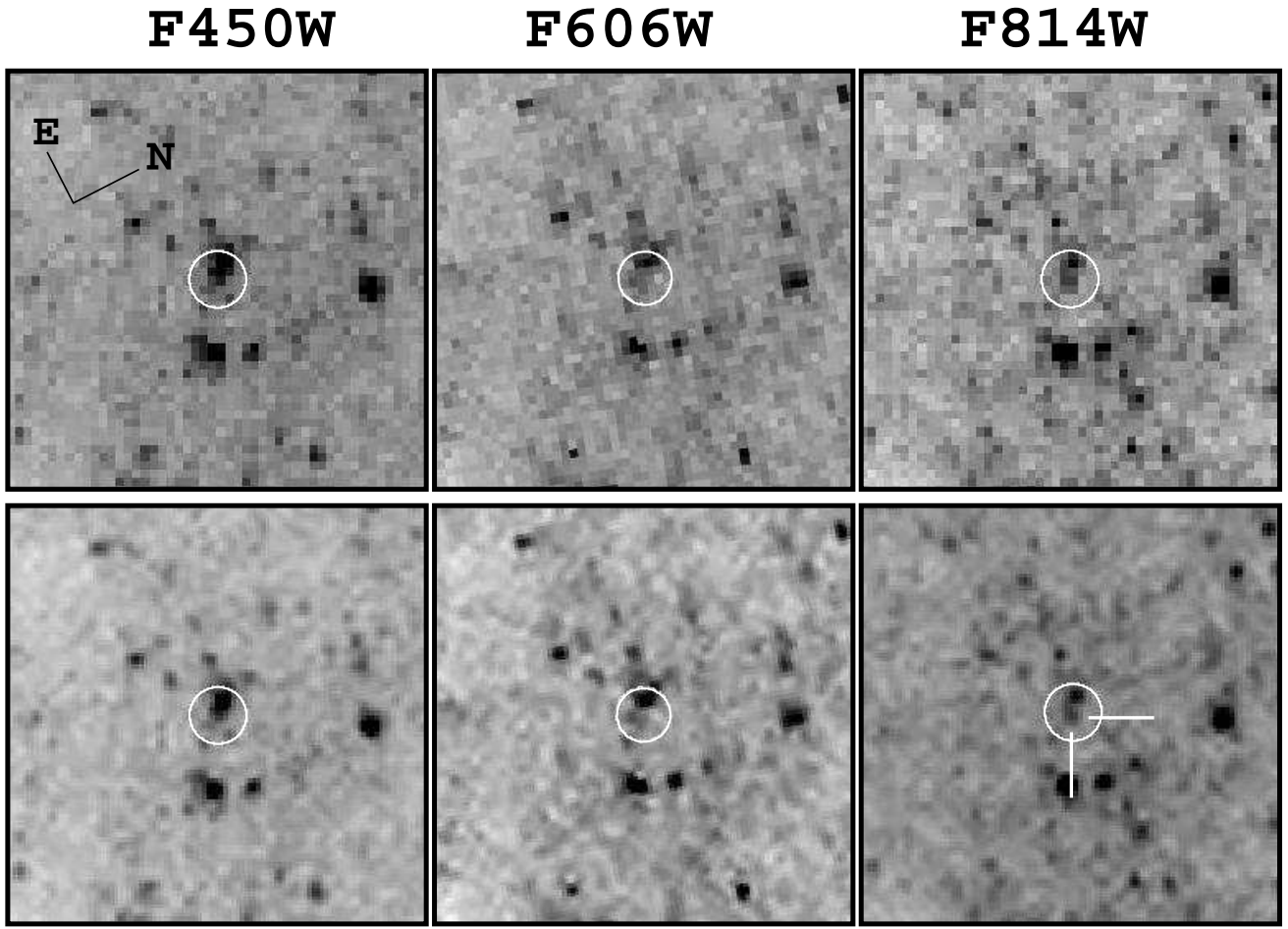}{6.4in}{0}{70}{70}{00}{60}}
\caption[] {
A $5\arcsec \times 5\arcsec$ close-up of the SN 2006ov environment in the
$HST$/WFPC2 F450W, F606W, and F814W images. The white circles mark 
20 times the $1\sigma$ uncertainty of the astrometric registration.
The F606W image is rotated to match the orientation of the F450W and F814W 
images. The upper panel shows the images in 
the original WFPC2 resolution (0$\farcs$1/pixel), while the lower panel
resamples the data to a resolution of 0$\farcs$05/pixel to bring out more  
details. The candidate progenitor is marked by a cross hair in the
resampled F814W image. Photometry of the stars was performed on the original images.
}
\label{8}
\end{figure*}

\begin{figure*}
{\plotfiddle{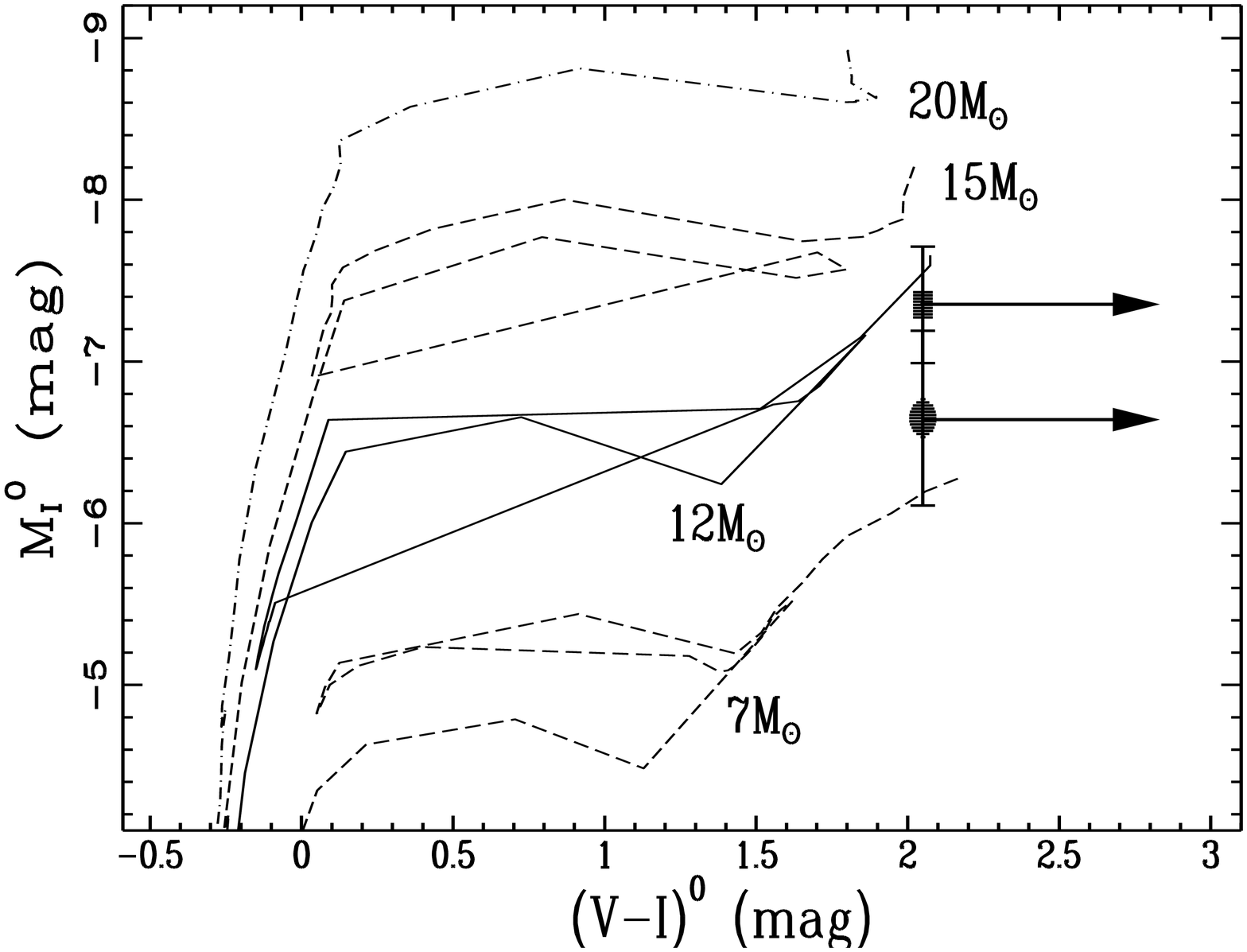}{6.4in}{0}{70}{70}{-40}{90}}
\caption[] {The $(V-I)^0$ vs. $M^0_I$ color-magnitude diagram for the 
progenitor of
SN 2006my. The filled square represents the data with an adopted distance
modulus of $\mu = 31.8 \pm 0.3$ mag for SN 2006my, while the filled circle
represents the data with $\mu = 31.1 \pm 0.5$ mag.  Also shown are model
stellar evolution tracks for a range of masses from Lejeune \& 
Schaerer (2001), with enhanced mass loss for the most massive stars and 
a metallicity of $Z = 0.008$. 
} 
\label{9}
\end{figure*}

\begin{figure*}
{\plotfiddle{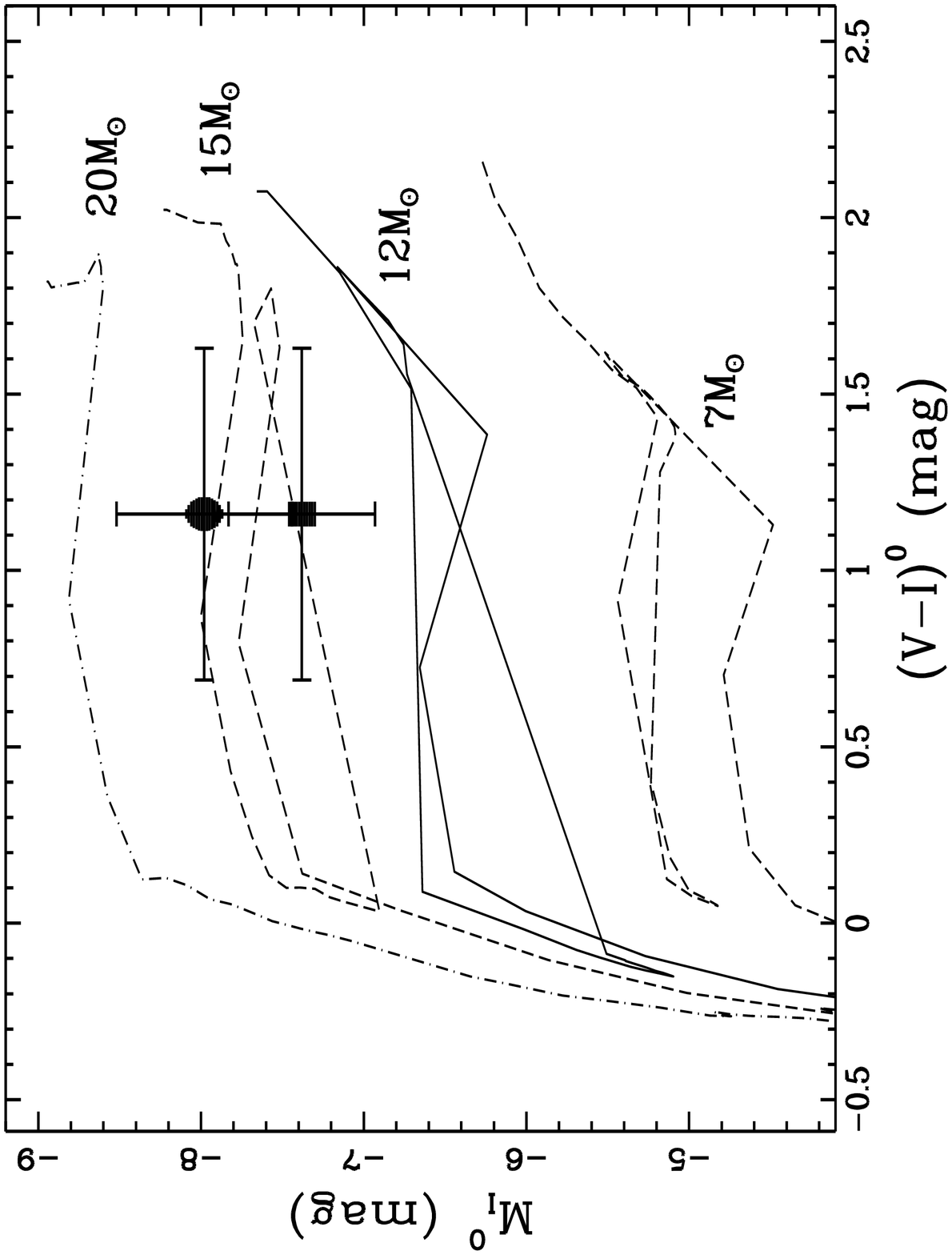}{6.4in}{-90}{70}{70}{-40}{550}}
\caption[] {The $(V-I)^0$ vs. $M^0_I$ color-magnitude diagram for the 
progenitor of
SN 2006ov. The filled square represents the data with an adopted distance
modulus of $\mu = 30.5 \pm 0.4$ mag for SN 2006ov, while the filled circle
represents the data with $\mu = 31.1 \pm 0.5$ mag.  Also shown are model
stellar evolution tracks for a range of masses from Lejeune \& 
Schaerer (2001), with enhanced mass loss for the most massive stars and 
a metallicity of $Z = 0.008$. 
} 
\label{10}
\end{figure*}

\begin{figure*}
{\plotfiddle{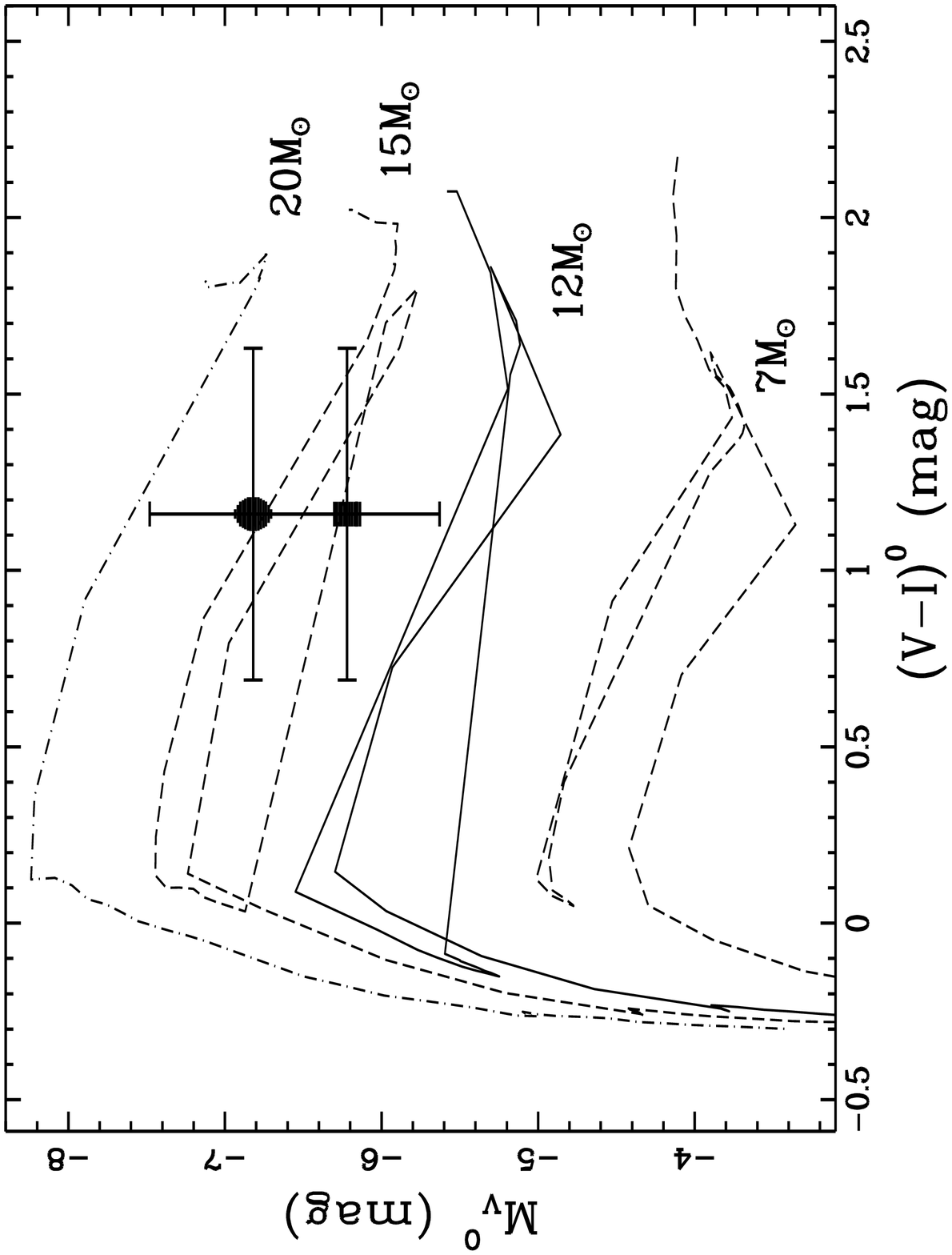}{6.4in}{-90}{70}{70}{-40}{550}}
\caption[] {The $(V-I)^0$ vs. $M^0_V$ color-magnitude diagram for the 
progenitor of
SN 2006ov. The filled square represents the data with an adopted distance
modulus of $\mu = 30.5 \pm 0.4$ mag for SN 2006ov, while the filled circle
represents the data with $\mu = 31.1 \pm 0.5$ mag.  Also shown are model
stellar evolution tracks for a range of masses from Lejeune \& 
Schaerer (2001), with enhanced mass loss for the most massive stars and 
a metallicity of $Z = 0.008$. 
} 
\label{11}
\end{figure*}

\newpage

\renewcommand{\arraystretch}{0.75}

\begin{deluxetable}{llrllccl}
\tablecaption{Journal of Spectroscopic Observations of SNe 2006my and 2006ov}
\tablehead{
\colhead{SN}&\colhead{UT Date} & \colhead{$t$(disc.)\tablenotemark{a}}&\colhead{Tel./Instrument}
&\colhead{Range~(\AA)\tablenotemark{b}}& \colhead{Air.\tablenotemark{c}} &
\colhead{Slit} &\colhead{Exp. (s)}  
}
\startdata
2006my& 2006-11-23 & 15 &Keck I 10-m/LRIS&3086--9250  &1.55&1$\farcs$0 & 200 \\
2006my& 2006-12-01 & 23 &Lick 3-m/Kast&3310--10400 &1.25&2$\farcs$0 & 1500 \\
2006ov& 2006-12-01 &  7 &Lick 3-m/Kast&3310--10400 &1.48&2$\farcs$0 & 1500 \\
\enddata
\tablenotetext{a}{Time in days since discovery.  }
\tablenotetext{b}{Observed wavelength range of spectrum. }
\tablenotetext{c}{Average airmass of observations.}
\end{deluxetable}

\begin{deluxetable}{ccccccr}
\tablecaption{{\it HST}/WFPC2 Data for NGC 4651 Pre-SN 2006my}
\tablehead{
\colhead{Dataset} & \colhead{UT Date} &\colhead{Exp. (s)}
&\colhead{Filter} &\colhead{Prop. ID}
}
\startdata
u2dt0901t & 1994 May 20 &  60 & F555W & 5375 \\
u2dt0903t & 1994 May 20 & 300 & F555W &5375\\
u2dt0903t & 1994 May 20 & 300 & F555W &5375\\
u2dt0904t & 1994 May 20 &  60 & F814W &5375\\
u2dt0905t & 1994 May 20 & 300 & F814W &5375\\
u2dt0906t & 1994 May 20 & 300 & F814W &5375\\
u2ex0f01t\tablenotemark{a} & 1995 Mar  4 & 900.0 & F218W &5419\\
u2ex0f02t\tablenotemark{a} & 1995 Mar  4 & 900.0 & F218W &5419\\
u2ex0f03t\tablenotemark{a} & 1995 Mar  4 & 300.0 & F547M &5419\\
\enddata
\tablenotetext {a}{The site of SN 2006my was not imaged.}
\end{deluxetable}

\begin{deluxetable}{ccccccr}
\tablecaption{{\it HST}/WFPC2 Data for M61 Pre-SN 2006ov}
\tablehead{
\colhead{Dataset} & \colhead{UT Date} &\colhead{Exp. (s)}
&\colhead{Filter} &\colhead{Prop. ID}
}
\startdata
u29r4d01t & 1994 Jun  6 &  80 & F606W & 5446 \\
u29r4d02t & 1994 Jun  6 &  80 & F606W & 5446 \\
u33z0801t\tablenotemark{a} & 1996 Mar 15 & 600 & F218W & 6358 \\
u33z0802t\tablenotemark{a} & 1996 Mar 15 & 600 & F218W & 6358 \\
u6ea6201r & 2001 Jul 26 & 230 & F450W & 9042 \\
u6ea6202r & 2001 Jul 26 & 230 & F450W & 9042 \\
u6ea6203r & 2001 Jul 26 & 230 & F814W & 9042 \\
u6ea6204r & 2001 Jul 26 & 230 & F814W & 9042 \\
\enddata
\tablenotetext {a}{The site of SN 2006ov was not imaged.}
\end{deluxetable}

\begin{deluxetable}{llllll}
\tablecaption{Summary of CFHT MegaCam Observations}
\tablehead{
\colhead{SN}&\colhead{UT Date }&\colhead{Filter}&
\colhead{Exp. (s)}&\colhead{Pixel Scale}&
\colhead{Seeing}
}
\startdata
2006my &2006 Nov 19 & $r\arcmin$ & 60 & $0{\farcs}185$ & $0{\farcs}8$ \\
2006my &2006 Dec 24 & $r\arcmin$ & 60$\times$5 & $0{\farcs}185$ & $0{\farcs}6$ \\
2006ov &2006 Nov 27 & $r\arcmin$ & 48$\times$5 & $0{\farcs}185$ & $1{\farcs}0$ \\
\enddata    
\end{deluxetable}

\begin{deluxetable}{ccccccr}
\tablecaption{{\it HST}/ACS Observations of SN 2006ov}
\tablehead{
\colhead{Dataset\tablenotemark{a}} & \colhead{UT Date} &\colhead{Exp. (s)}
&\colhead{Filter} &\colhead{Prop. ID}
}
\startdata
 j9nw50011& 2006 Dec 12 & 420 & F435W & 10877 \\
 j9nw50021& 2006 Dec 12 & 180 & F625W & 10877 \\
\enddata
\tablenotetext {a}{The site of SN 2006ov was not imaged.}
\end{deluxetable}

\begin{deluxetable}{lllllllll}
\tablecaption{HSTphot Photometry for the Progenitors of SNe 2006my and 2006ov}
\tablehead{ 
\colhead{SN}&\colhead{X(p)\tablenotemark{a}}&\colhead{Y(p)\tablenotemark{a}}&
\colhead{X(m)\tablenotemark{b}}&\colhead{Y(m)\tablenotemark{b}}&\colhead{S/N} & \colhead{Mag1\tablenotemark{c}} &
\colhead{Mag2\tablenotemark{c}}&\colhead{Lmag\tablenotemark{d}}
}
\startdata
2006my&410.11&158.31& 409.72& 158.13& 5.6& F814W=24.47(20)& $I$=24.45(20) &$I<$25.5 \\
2006my&410.11&158.31& $-$& $-$ & $-$& F555W$<$26.5& $V<$26.5&$V$$<$26.5  \\
2006ov&571.12&235.72& 571.12\tablenotemark{e}& 235.72\tablenotemark{e}& 6.1& F814W=23.19(18)& $I$=23.08(18)&$I<$24.6 \\
2006ov&226.94&266.84& 226.94\tablenotemark{e}& 266.84\tablenotemark{e}& 2.2& F606W=24.07(50)& $V$=24.24(50)&$V<$25.3 \\
2006ov&571.12&235.72& 571.12\tablenotemark{e}& 235.72\tablenotemark{e}& 6.1\tablenotemark{f}& F450W=23.51(18)\tablenotemark{f}& $B$=23.40(18)\tablenotemark{f}&$B<$25.3 \\
\enddata    
\tablenotetext{a}{The coordinates as predicted from the astrometric solutions.}
\tablenotetext{b}{The coordinates as measured in HSTphot.}
\tablenotetext{c}{Uncertainties in the last two digits for the magnitudes are indicated in parentheses.}
\tablenotetext{d}{The 3$\sigma$ limiting magnitude. See text for details. }
\tablenotetext{e}{The coordinates as enforced in HSTphot. See text for details.}
\tablenotetext{f}{This detection is likely caused by another source and not the 
progenitor of SN 2006ov. See text for details.} 
\end{deluxetable}

\begin{deluxetable}{lllll}
\tablecaption{Masses and Mass Limits for the Progenitors of Core-Collapse SNe}
\tablehead{
\colhead{SN}&\colhead{SN type} &\colhead{Progenitor mass} 
&\colhead{Progenitor mass limit} &\colhead{Ref\tablenotemark{a}}
}
\startdata
1987A  & II-peculiar&  $\sim$20 $M_\odot$&                 &1,2\\
1993J  & IIb        &  $\sim$17 $M_\odot$&                 &3,4\\
1999ev & II-P       &  15--18 $M_\odot$&                   &5\\
2003gd & II-P       &   6--12 $M_\odot$&                   &6,7\\
2004A  & II-P       &   7--12 $M_\odot$&                   &8  \\
2004et & II-P       &  13--20 $M_\odot$&                   &9\\
2005cs & II-P       &   7--13 $M_\odot$&                   &10,11\\
2006my & II-P       &   7--15 $M_\odot$&                   &this paper\\
2006ov & II-P       &  12--20 $M_\odot$&                   &this paper\\
2005gl & IIn? II-L? &                  & 40--80 $M_\odot$ LBV? CSC?  &12\\
2004dj & II-P       &                  & $\sim 12-15 M_\odot$? $> 20 M_\odot$? (in CSC)&13,14,15\\
1999em & II-P       &                  &  $\ale 15 M_\odot$    &16\\
1999gi & II-P       &                  &  $\ale 12$--$20 M_\odot$&17\\
2001du & II-P       &                  &  $\ale 9$--$21 M_\odot$ &18,16\\
1999an & II-P       &                  &  $\ale 20 M_\odot$    &19\\
1999br & II-P       &                  &  $\ale 12 M_\odot$    &19\\
2000ds & Ib/c       &                  & $\ale 7 M_\odot$ RSG? W-R? &19\\
2000ew & Ib/c       &                  &  low-mass RSG; W-R?        &19\\
2001B  & Ib/c       &                  & $\ale 25 M_\odot$ RSG? W-R?&19\\
2004gt & Ib/c       &                  & $\sim 20$--$40 M_\odot$? W-R?&20,21\\
\enddata
\tablenotetext{a}{ References:
1. Gilmozzi \etal 1987; 2. Sonneborn \etal 1987; 3. Aldering \etal 1994; 4. Van Dyk et al. 2002; 5. Maund \& Smartt 2005; 6. Van Dyk \etal 2003c; 7. Smartt \etal 2004; 8. Hendry et al. 2006; 9. Li \etal 2005a; 10. Li et al. 2006; 11. Maund et al. 2005a; 12. Gal-Yam et al. 2007;
13. Wang \etal 2005; 14. Ma\'iz-Apell\'aniz \etal 2004; 15. Vink\'o \etal 2006; 16. Smartt \etal 2003; 17. Leonard et al. 2002a; 18. Van Dyk \etal 2003b; 19. Maund \& Smartt 2005; 20. Gal-Yam et al. 2005; 21. Maund \etal 2005b.
}
\end{deluxetable}
\end{document}